# Preliminary Guideline for Creating Boundary Artefacts in Software Engineering

Raquel Ouriques[a], Fabian Fagerholm[b], Daniel Mendez[a,c], Tony Gorschek[a,c], Baldvin Gislason Bern[d]

[a]*Blekinge Institute of Technology, Karlskrona, 37179, Sweden*
[b]*Aalto University, Espoo, FI-00076, Finland*
[c]*fortiss, Munich, 80805, Germany*
[d]*Axis Communications, Lund, 22369, Sweden*

**Abstract**

**Context**: Software development benefits from having Boundary Artefacts (BAs), as a single artefact can supply stakeholders with different boundaries, facilitating collaboration among social worlds. When those artefacts display inconsistencies, such as incorrect information, the practitioners have decreased trust in the BA. As trust is an essential factor guiding the utilisation of BAs in software projects, it is necessary to understand which principles should be observed when creating them.

**Objective**: This study aimed at develop and validate a preliminary guideline support the creation of trustworthy BAs.

**Method**: We followed a multi-step approach. We developed our guideline through a literature review and previous results from our case study. Second, we submitted the guideline for an expert evaluation via two workshops and a survey. At last, we adjusted our guideline by incorporating the feedback obtained during the workshops.

**Results**: We grouped the principles collected from a literature review into three categories. The first category (Scope) focuses on the scope, displaying principles referring to defining each boundary's target audience, needs, and terminology. The second category (Structure) relates to how the artefact's content is structured to meet stakeholders' needs. The third (Management) refers to principles that can guide the establishment of practices to manage the artefact throughout time. The expert validation revealed that the principles contribute to creating trustworthy BAs at different levels. Also, the relevance of the guideline and its usefulness.

**Conclusions**: The guideline strengthen BA traits such as shared understanding, plasticity and ability to transfer. Practitioners can utilise the guideline to guide the creation or even evaluate current practices for existing BAs.

# 1. Introduction

Boundary artefacts (BA) are objects crossing the boundaries of social worlds, carrying several meanings that meet the information needs of different stakeholders [1]. Software development utilises BAs to a large extent and for different purposes, including requirements specifications, architectural descriptions, and project planning documents [2, 3].

Software development benefits from BAs by supplying different teams with condensed information in several formats. Also, these artefacts provide the means for keeping and carrying relevant information helping coordination among several stakeholders. For example, software requirements specifications are used by, and thus support, different roles such as architects, project managers, and stakeholders at customer side, by conveying information that has different purposes for each role across each of the boundaries.

While BAs strengthen collaboration by providing timely and correct content throughout development projects, inconsistencies in BAs can affect how practitioners perceive and utilise them [4, 5].

The organisation of BAs affects the trust that practitioners deposit in them and also influences their behaviour when utilising those artefacts [4]. When practitioners do not feel confident about BAs, their level of trust decreases. Besides the effects on utilisation, several negative implications for software projects can emerge, such as a setback in the development process, uncertainty about the availability of the information, time wasted looking for information and wrong execution of tasks [4].

Decreased trust levels can also influence practitioners not to use the BAs and re-create their solutions, causing frustration. Such inefficiency in managing companies' knowledge through artefacts can result in increased costs and time waste [6].

As trust is an essential factor guiding the utilisation of BAs in software projects, there is a need to understand which principles should be observed when creating such artefacts and to what extent these principles contribute to creating trustworthy artefacts [4].

Although there seems to be a common understanding of the vast production and use of artefacts in software engineering, to the best of our knowledge, the community still lacks necessary guidelines required to support an effective creation and management of BAs. We contribute to filling this gap by investigating principles that support the creation of trustworthy BAs.

In this manuscript, we report on the following contributions:

- We develop preliminary guideline focusing on the creation of BAs.

- We conduct an expert validation with a company partner to validate our proposed guideline and implement their feedback.

- We collect the participants' perceptions on how the proposed guideline contribute to creating trustworthy BAs and their usefulness.

- We provide supplementary material for transparency and potential re-application of the guideline in other contexts.

This manuscript is organised as follows: Section 2 presents a brief background to our study and related work. Section 3 describes our research methodology. Section 4 presents the developed guideline and the results of the expert validation. Section 5 provides a discussion of the results and implications that our current findings have for future research. In Section 6, we discuss the lessons learned after our validation. Section 7 describes the threats to the validity of our study. Lastly, in Section 8, we present our concluding remarks.

## 2. Background and Related Work

In this section, we introduce the basic notion of BAs and how trust relates to inanimate software artefacts as the background of our study. We conclude by discussing related work.

### 2.1. Boundary Artefacts and Trust

The term boundary artefact (often referred also as boundary object), was originally coined in a study by Star and Griesemer in the Museum of Vertebrate Zoology [1]. We adopt the term boundary artefact here due to its closeness to the software engineering. The authors refer to the term as an object having enough flexibility to adapt to local needs, providing shared understanding among different users while keeping its identity as they cross many boundaries. In their work, the authors offer the example of a university administrator world who is often involved in contracts and grants, performing a particular set of tasks and dealing with different audiences. At some point, the administrator's world intersects with the world of a naturalist, in which part of the job is to collect specimens for a natural history museum. When this happens, the boundary object will minimize the difficulties with communication and cooperation by providing content that can satisfy their needs even though having different meaning in different worlds.

Boundaries are not always synonyms with a limit in organisations. Instead, they are mostly as permeable. That is, they are not stoppers of information in an organisation, but there is a transition where people on one side of the boundary see information from their perspective, while people on the other side see it from another perspective – and they might be looking at the same information or working towards the same goals. For example, a use case developed to analyse requirements will be utilised in many other software development activities such as project organisation and testing. Every time the artefact crosses a boundary, its purpose can change depending on the people's needs.

Boundary artefacts have a few characteristics that help distinguish them from other types of artefacts. They can provide shared meaning by satisfying people's needs in different social worlds. To be able to do that, they are plastic enough to meet local needs and keep a common identity when crossing boundaries. When in common use, they are weakly structured but become strong in individual use due to their ability to meet local needs. Their representation can be abstract or concrete [1].

Software development benefits from having BAs as a single artefact can supply stakeholders in different boundaries, especially in geographically distributed environments. However, when lacking proper management, these artefacts can be anywhere from being overloaded with meanings that render it difficult to find relevant information to the absence of enough details. In such a situation, BAs can lose their value to stakeholders and decrease their efficacy [7, 8, 9].

Intergroup politics around BAs can affect how stakeholders utilise them. They are not always politically neutral, and their subjective dimension often makes them dependent on trustworthiness [10, 11]. Trust plays a relevant role in driving the positive behaviour of stakeholders towards BAs.In inanimate artefacts, trust has been explored under the premise that people perceive inanimate software artefacts as possessing human attributes [12, 13].

There is no consensus about a definition of the term, but for a general understanding, trust refers to how individuals deliberately rely on others [14]. Due to its abstract characteristic, it is complex to assess. To address this complexity, trusting beliefs have been developed to help describe the favourable factors that make individuals trustworthy, such as benevolence, integrity, predictability, and competence [14, 15, 16].

An extensive study on information technology investigated which trust beliefs best describe trust in inanimate software artefacts [17]. The authors conducted a literature review and identified the following trusting beliefs: Reliability - Perception that the artefact provides accurate content. Predictability - Confidence that the arte-fact content is always provisioned as requested. Provide the required functionality - perform as needed for the task environment.

Trust in inanimate software artefacts is little examined in software engineering [4], especially principles that contribute to creating trustworthy BAs. Principles are dispersed and not examined from the trust perspective.

*2.2. Related work*

Few studies provide solutions or practices targeting BA in software engineering. Radhika et al. [18] investigate BAs' roles in addressing requirements engineering challenges when having multiple stakeholders in the product family development context. Through a multiple case study, they provide insights on utilising BAs to improve the quality of requirements engineering processes and suggestions for managing such artefacts.

Focusing on the automotive industry, Wohlrab at al. [8] created practical guideline for managing systems engineering artefacts. They analysed each type of artefact's challenges and provided solutions that were further packed into contextualised guideline.

Loggem and Veer argue [19] favouring a documentation-centred approach to support internal and external communication (focusing on users). According to the authors, the BAs potentially solve two main issues, heterogeneous mental models within teams and the peripheral role of the user concerning the teams.

We have previously investigate how differences in trust level can affect stakeholders' behaviour towards BAs [4]. When the content of the BAs does not meet stakeholders' needs, several implications affect the software development project. To minimise the implications, the authors offer potential solutions.

Although the referred studies offer suggestions for practices regarding the creation and content management of BAs, the practices are particular to specific cases. We believe this is the first work that gathers generic principles into preliminary guideline that support the creation of BAs in software engineering. In the interest of advancing research into trust in inanimate software artefacts, this study adds to our understanding of how the guideline' principles can help create trustworthy BAs.

## 3. Research Methodology

This study aimed at developing and validating preliminary guideline for creating boundary artefacts in software engineering.

Our motivation for developing such guideline originates from previous research gaps [4] and discussions with a company partner collaborating in this study. We then set the following Research Questions (RQ):

- RQ1: What principles should guide the creation of boundary artefacts?
- RQ2: To what extent can a BA creation guideline be used to create trustworthy boundary artefacts?
- RQ3: To what extent do practitioners perceive the guideline to be effective for creating boundary artefacts?

To answer those questions, we utilised a mixed methods approach performed in three steps (see Fig. 1). First, we developed our guideline through a literature review and previous results from our case study [4]. Second, we submitted the guideline for an expert evaluation via two workshops and a survey. At last, we adjusted our guideline by incorporating the feedback obtained during the workshops. We detail the three steps in the following subsections.

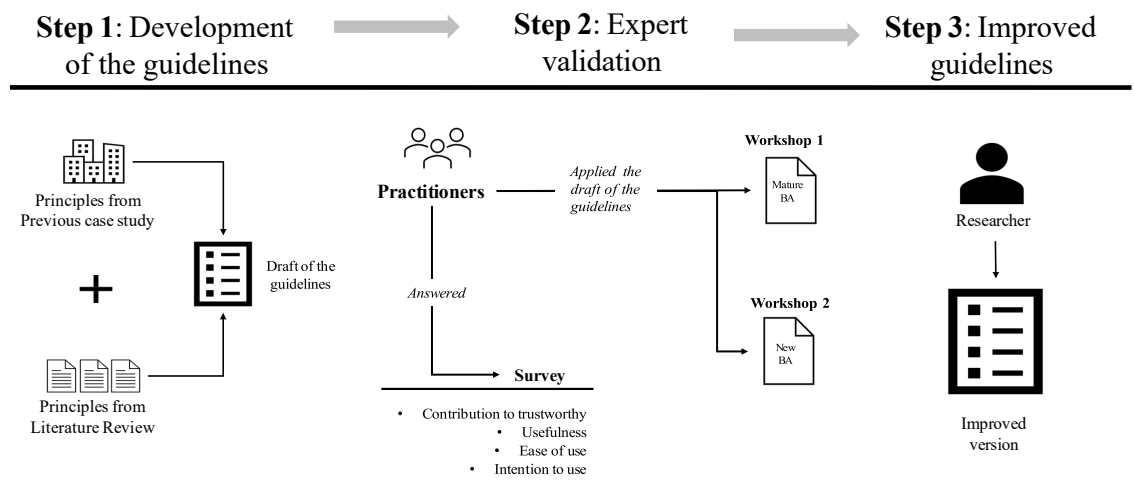

Figure 1: Research approach followed for developing the guideline.

### 3.1. **Step 1** - *Development of the guideline*

In the first step (see step 1 in Fig.1), we collected input from two primary sources to develop the first version of the guideline: a synthesis of the literature and previous case study results [4].

We performed four searches in the Scopus database with distinct search strings (see Table 1). The goal was to find relevant studies that reported guideline, frameworks, checklists, and practices utilised to create and manage BAs. We did not limit

Table 1: Search strings utilised for finding relevant papers.

| Search Strings | Date | Hits | Number of Selected Papers |
|---|---|---|---|
| 1 - (("boundary artefact" OR "boundary object") AND (feedback OR "feedback loop")) | 10/2022 | 27 | 1 |
| 2 - ("Designing boundary artefact" OR "designing boundary objects") | 10/2022 | 2 | 0 |
| 3 - ("creating boundary artefact" OR "creating boundary objects") | 10/2022 | 4 | 2 |
| 4 - (("boundary artefact" OR "boundary objects") AND ("software development" OR "software engineering")) | 10/2022 | 65 | 9 |

ourselves to the software engineering literature. We made the search as broad as possible.

Please note that our goal was not to perform a rigorous systematic literature review as widely established by Kitchenham et al.[20]. We selected the studies by screening their abstracts and introduction. We provide the list of selected studies and principles collected from each in the supplemental material.

We also added principles extracted from our previous case study [4]. Later, we aggregated similar guides into three categories, detailed in subsection 4.2.

## 3.2. **Step 2** - *Expert validation*

In our study, we validated our guideline with experts and collected feedback for further improvement (see step 2 in Fig.1). We also collected their perception of the guideline regarding the following rubrics (details of the dynamics of the workshop can be taken from the supplementary material):

- **Appropriateness criteria of the guideline** - how practitioners perceive the guideline to be appropriate for creating BAs.

- **Contribution to trustworthiness** - The degree to which practitioners perceive the guideline to contribute to creating trustworthy BAs. Here, we utilise the trusting beliefs proposed by Lansing and Sunyaev [17]: reliability, Functionality, and Predictability.

- **Ease of use, Usefulness, and Intention to use** – We followed the theoretical model for validating information system design methods [21]. We evaluate the guideline concerning the degree to which practitioners believe the guideline would be: free of effort (ease of use) and the degree to which they believe that the guideline will be effective in achieving their intended objective (usefulness). Lastly, the degree to which practitioners intend to use the proposed guideline.

We performed our workshops at Axis Communications, which provides network solutions in video surveillance, access control, intercom, and audio systems. Axis has over 4,000 employees in over 50 countries, and its headquarters is in Lund, Sweden, where the validation activities were centred.

The workshops lasted around 90 minutes each. A total of six practitioners applied the guideline to two BAs in different maturity stages. In both workshops, we recorded the audio for later collecting their feedback.

*Workshop 1*

Five practitioners participated in the first workshop. They utilise the BA1 daily, creating or applying its content as input for their tasks. BA1 has been largely utilised in the company for many years. In this workshop, the practitioners examined the BA against our guideline to verify whether they matched the current practices, the status of each principle, and how important they were for the future.

**Description of the BA1**: It presents the core list of features of all the products in a software product line of the company (covering over 200 products). This content is stored in an XML file with the developed features for the firmware used in the company's products, whether new, modified or deprecated.

*Workshop 2*

In the second workshop, three practitioners from workshop 1 also joined due to their close connection to BA2 - which was initially planned to complement BA1. In this workshop, practitioners would analyse if the principles would be relevant in creating BA2 and briefly discuss if and how they would implement them.

**Description of the BA2**: It describes the logical relationship between facts that are expected to be true, e.g. a rule could be "if a product in the software product line has Feature A, then the product shall also have Feature B". The rules are described in YAML format and can be used to automatically detect whether products in the product line follow the rules.

*Survey*

At the end of the second workshop, the first author sent a survey (see table 2) to all participants in both workshops. In that survey, they would answer how they perceived each principle contributing to the trust factors' reliability, predictability and functionality. They also evaluate the guideline' usefulness, ease of use, and intent to use. A total of five practitioners answered the survey. The data analysis from the survey was done through descriptive statistics, focusing on the frequency of the responses. We also utilised stacked bar charts for visualisation (see subsection 4.3).

| RUBRICS | QUESTIONS |
| --- | --- |
| Predictability | To what extent do these principles contribute to the content always being available? |
| Reliability | To what extent do these principles contribute to the correctness of content in an artefact? |
| Functionality | To what extent do these principles contribute to the proper functionality of the artefact? |
| Ease of use | Please indicate the degree you believe the guideline are easy to follow. |
| Usefulness | Please indicate the degree you believe the guideline are useful. |
| Intention to use | How likely are you intended to use the guideline? Under which conditions would you use it? |

Table 2: Survey Questions

### 3.3. **Step 3** - *guideline Improvement*

We collected feedback on the guideline during both workshops and adjusted its design. The feedback focused on the relevance of each principle to the guideline, general structure, and clarity of the content. After both workshops, we improved our guideline, which we detail in subsection 4.2.

Regarding the presentation of the guideline and the placement of the principles, the guideline went through one revision after they were first planned. The first version had four categories. The additional category (shared understanding) had one principle (Define an approach to deal with different interpretations of terminology). However, for simplicity, after receiving feedback from practitioners, we relocated the principle to the category Structure justified by its close connection to the theme. When designing the BA, identifying stakeholders' needs already provides hints on how terminology will differ when crossing borders.

We also moved principle 3d. Agree upon a balanced formality that was initially in the Structure category as it was more coherent with the scope of the category Management.

Changes relating to interpretation and misinterpretation focused on offering better and clear descriptions of the principles. For example, establishing ownership descriptions had too many academic terms, which made it difficult for practitioners to understand. They also asked for standardization of the writing (using active verbs to start) and the definition of each category.

*3.4. Ethical concerns*

Our ethical concerns approach focused on three main aspects: data collection, data analysis, and handling the company's confidential information. We designed and executed our study per the guideline for ethical research provided by the Swedish Research Council [22].

In the data collection phase, we carefully elaborated the questions to guide the workshop not to raise intense emotions or provoke emotional harm [23] (see also our supplementary material). We also disclosed the study's goal and how and when they could remove the data they provided. Also, we explained how the data would be utilised and anonymised through informed consent (see supplementary material).

In the data analysis phase, we anonymised quotes from practitioners and confidential information about the BAs utilised in the workshops. When reporting the results, we carefully paid attention to not stigmatise or harming specific populations. A company employee also checked the handling of confidential information in our manuscript before submission.

## 4. Results

In this section, we report on the results of our study. Besides explaining how we gathered the principles (see subsection 4.1), we divided it into two major subsections, which report the latest version of the guideline (RQ1) and the results of the expert validation (RQ2 and RQ3). We also display an example of the application of the guideline at Axis (see subsection 4.4).

*4.1. The organisation of the guideline*

The developed categories emerged from our analysis as we collected the principles from the literature. When collecting them, we observed similarities in what they referred to, which made sense in gathering them into categories. For example, principles 2a and 2b (Decide on the format of the artefact and Establish the means of spreading the artefact) strongly connect with how people think about arranging the content, its layout and which channels work best for spreading it. Thus, for principles involving content manipulation, we grouped them into the category structure.

*4.2. The BA-Guide: Preliminary guideline for creating boundary artefacts in software engineering*

The BA-Guide (see Table 3) provides ten principles supporting the creation of boundary artefacts in software engineering. We describe the principles and supplement them with examples of implementation. We organise the guideline into scope, structure, and management categories.

*1 - Scope*

The scope category displays the principles referring to the definition of the target audience in each boundary, their needs regarding the artefact, and how different terminology is held across boundaries.

*2 - Structure*

The principles in this category relate to how the artefact's content is structured to meet stakeholders' needs regarding format, level of detail, and formality.

*3 - Management*

The category refers to principles that can guide the establishment of practices to manage the artefact throughout time.

Table 3: The BA-Guide: Preliminary guideline for creating boundary artefacts in software engineering

| CATEGORY | PRINCIPLES | DESCRIPTION |
| --- | --- | --- |
| **1. SCOPE** | 1a. Identify the stakeholders and the boundaries | Identification of the stakeholders that will utilise the artefact's content and their information needs. They affect the usage of the content and how the content is created [24, 25, 4]. Boundaries are usually blurry and often do not relate to a limit or edge but a shared space. They can be teams, departments, units, a group of teams, or, more commonly, organisational domains. It is important to identify them, as the terminology can have several meanings when crossing boundaries. After the workshop with the practitioners, we observed that it is easier to start identifying all the stakeholders that will use the BA and, later, draw the boundaries among them. |
| | 1b. Define an approach to deal with different interpreta-tions of termi-nology | Refers to how the terminology will be translated through several boundaries [4, 25, 18]. Will the artefact have a glossary of terms or provide commonality and differences in processes across the boundaries? Will it have additional materials that help stakeholders interpret the terminology? |
| **2. STRUCTURE** | 2a. Decide on the format of the artefact | Refers to how the content is organised and packed in the artefact [25]. The format in which the content is delivered affects how stakeholders will utilise it. Which format is the best? Does it attend to all stakeholders' needs? XML files? Pictures? Which one attends to all stakeholders' needs? |
| | 2b. Establish the means of spreading the artefact | Refers to how the artefact will cross the boundaries and reach the stakeholders [25]. People involved in creating such artefacts should pay attention to how the audience will engage with the content and plan for a simple interaction of finding and utilising the content. As an example of the means, we could consider using Whiteboards, Git repository, Webserver, Confluence pages, etc. |
| | 2c. Decide an appropriate level of content detail | Refers to the level of detail of the content. Enough details in the content are recommended, so the stakeholders find the exact information being searched. If there is too much content, the artefact can be polluted with unnecessary details and decrease its utilisation. On the contrary, the stakeholders will create workarounds that can produce scattered information [8]. |
| | 2d. Accommodate additional and experimental content | Refers to the flexibility of the artefact regarding testing new content and increasing its amount. Software development is a dynamic activity that usually welcomes changes while developing and producing experimental content [19, 4]. Does the artefact allow adding additional content (flexibility) to supply any particular need? Thinking about the future, does the actual artefact allow the addition of experimental content that can be incorporated or removed at any time? |
| **3. MANAGEMENT** | 3a. Set up an evaluation and feedback process | Refers to developing a periodic evaluation of the content of the artefact (if it fulfils stakeholders' needs) and a plan for receiving and integrating feedback [24, 25, 18, 4]. The evaluation planning depends on the utilisation of the artefact. Will it be mostly automated? Maybe analysing bug reports or accessing logs of a specific tool can support the evaluation. Questionnaires also work for any artefact. They are simple to create and disseminate. At the same time, you can collect both feedback and an assessment. |
| | 3b. Establish ownership of the artefact | Refers to the distribution of responsibility over the artefact content (or parts of it) and managing changes and improvements [25, 19]. Implementing this guide can reduce incorrect or missing content in the BA and should be determined as soon as possible so that inconsistencies can be found and fixed in the early stages. |
| | 3c. Agree upon a balanced formality | Refers to the level of formality applied to the management of artefacts [25]. How many people are needed to authorise the artefact's incremental changes? How many steps are needed to implement new content/feedback? The level of formality should satisfy stakeholders' needs without being burdensome to create and maintain. |
| | 3d. Handle scattered information | Refers to avoiding and monitoring where and why parts of the content are scattered [8]. The verification can be done during the feedback collection, checking if stakeholders' needs are satisfied. Making the information easy to find and collect minimises the risk of stakeholders forking the content to create other artefacts that suits them better. |

*4.3. Contribution of the guideline to creating trustworthy artefacts*

After conducting the workshops, we sent the survey to collect the practitioners' perceptions on how the guideline' principles contributed to creating trustworthy BAs (RQ2). We display the results in Fig. 2. We also show the practitioners' evaluation regarding the ease of use, usefulness and intention to use the guideline.

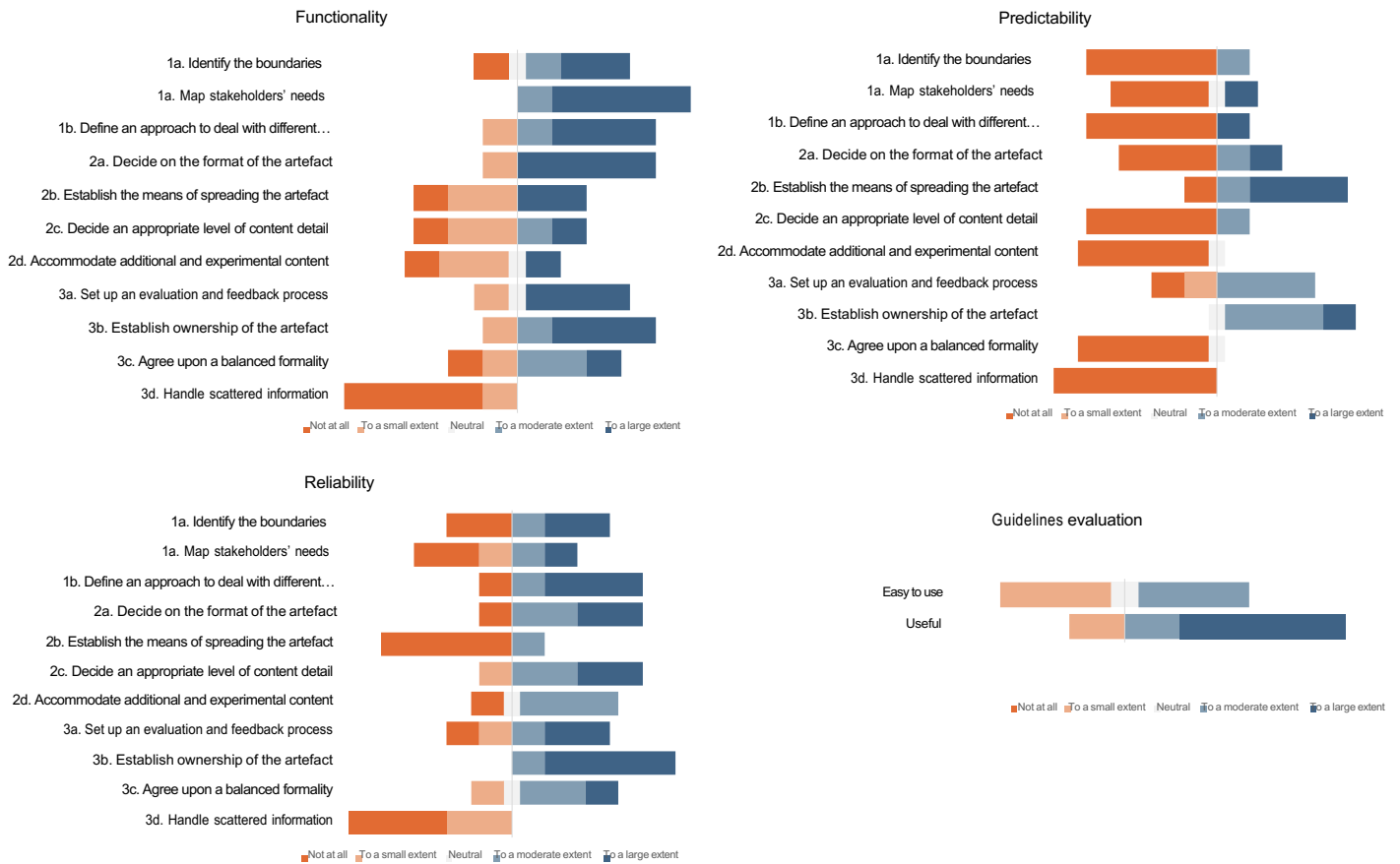

Figure 2: Contribution of the guideline to trust factors

We asked the practitioners how they perceived the guideline contributing to the BA performing as needed for the task environment (**Functionality**). In Figure 2, the practitioners perceive that seven principles contribute positively to functionality. The functionality is achieved mostly by the principles that outline the BA (category Scope) and establish practices to manage and keep (category Management) its construction.

When taking a closer look at principles 1a, 2a, and 1b, we observe that stakeholders expect that the content matches their needs and is displayed in a format that facilitates its usage. As the BA crosses boundaries, its functionality is enhanced by translating terminology to the different stakeholders.

The **Reliability** results are similar to the functionality, with a small, still positive, variation between moderate and large extent. However, principle 2b - Establish the means of spreading the artefact does not seem to contribute to the artefact providing accurate content, as pointed out by the practitioners. One reason for this result is that spreading the content occurs one step after the content is prepared (created or fixed). Therefore, it does not directly affect the accuracy of the content.

On the other hand, the practitioners consider principle 2b relevant for the artefact content is always provisioned as requested (**Predictability**). The other two principles are positively influencing the predictability of BAs. They are the existence of practices for evaluation and feedback implementation and the establishment of ownership, which can guarantee that the BA is continually managed.

The practitioners, on the whole, consider principle 3b - Establish ownership of the artefact as a positive and relevant principle that can contribute to the artefact being predictable, reliable and offering proper functionality.

In contrast to the other principles, principle 3c - handle scattered information, was not seen as contributing to any of the trust factors analysed during the workshops. During the discussions in the first workshop, the practitioners understood that if principle 1a (Map stakeholders' needs) is well-taken care of, scattered information won't probably happen. In the second workshop, they did not consider this principle when creating a BA simply because it did not exist yet, so scattered information was nonexistent.

In the final part of the survey, practitioners were asked about the guideline' ease of use, usefulness, and intention to use them. Figure 2 displays the results. There is no consensus among the practitioners that the guideline are easy to use. When crosschecking these results with the feedback received during both workshops, we believe that the practitioners had a hard time understanding the concept of BA and the complexity of the written language we utilised. By contrast, most practitioners believe that the guideline are useful to a large extent.

Concerning the intention to use, practitioners manifested a positive intention to apply the guideline in both artefacts that we utilised in the workshops. As one of the practitioners pointed out, "*one benefit of having some guideline like this is that if many boundary artefacts follow the same guideline or template, that will also make it easier for people to consume the boundary artefact. That makes me more convinced that we should use this even if it is not 100% perfect for everything; it will help people* *to align some of the different kinds of boundary artefacts we have and help people to work with them."*

Together these results provide important insights into the creation of trustworthy BAs. As the way the BA is structured and managed can influence stakeholders' behaviour, it is crucial to understand how we can drive this behaviour towards a positive attitude on which they can rely on them [4]. On the contrary, those knowledge resources will be underused and potentially increase the development costs due to the creation of complementary artefacts or wasting time trying to find the correct information.

*4.4. Example of application of the guideline*

After our last workshop, the practitioners developed a preliminary version of a template to guide schemas documentation creation (see Figure 3). The practitioners use schemas (i.e. common definitions of structured data in computer-readable format) to facilitate the automation of information flow.

A schema that has been used by the practitioners (BA1) had originally been poorly documented, and the hope is that better documentation, especially explicitly stating the Stakeholders, Terminology and Use cases, would help new schemas (such as BA2) to avoid some of the trust issues that were experienced with BA1. The schema document template was created to help creators of new schemas identify the minimum set of questions to be answered and documented for new schemas and align how different schemas are documented.

BA2's immediate stakeholders are people within the QA department, other R&D departments, and Product management. The practitioners also foresee that there will be interest in BA2 outside these departments, e.g. in Sales or Marketing.

## 5. Discussion

We followed a multi-step approach to develop and validate our preliminary guideline in an industry context. Our work expands on previous research on BAs by providing structured guideline gathered into three main categories (scope, structure, and management) that help practitioners create BAs in software engineering contexts. As the BA literature in software engineering contexts is scarce, this guideline is significant and unique because they offer an original perspective on essential principles that practitioners should observe upon the conception of a BA.

Further, in our guideline, we incorporated extendable practices applied to specific contexts such as product family development and the automotive industry [18, 8]. These were considered relevant for the context we utilised in this study, which we

**Schema Documentation Template**

**Owner**

Question(s) to be answered:

- *Who are the owners?*
- *Is it the same owner of the content and code?*

**Stakeholders**

Question(s) to be answered:

- *Who are the stakeholders?*

**Location**

Questions to be answered:

- *Where is the schema file located (e.g. git)?*
- *Is there more information to be found elsewhere (e.g. confluence space)?*

**Format**

Question(s) to be answered:

- *What is the format?*
- *Why are the elements/instances in the schema formatted as they are?*

**Terminology**

Question(s) to be answered:

- *What is the long name and short name of the schema?*
- *Are there terms that needs to be explained?*
- *What guidelines regarding the terminology will the schema follow?*
- *How/where will the terminology used in the schema be explained?*

**Use cases**

Question(s) to be answered:

- *What are the use cases (i.e. why is the schema needed)?*

Figure 3: Example - Schema Documentation Template

believe has a potential for generalisation of the principles as more applications of the guideline may occur.

The current version of our guideline evolved based on practitioners' feedback after conducting two workshops. Besides rearranging principles into categories (see subsection 3.3), the latest version differs from the first to meet practitioners' needs for clarity in describing each principle, especially regarding the concept of a boundary. During the workshops, they agreed with the principles' importance, although most were not implemented.

However, the concept of the boundary was difficult to understand, even when having a well-established BA. Attending to stakeholders' needs was tacitly made, but not all found the needed content. As we observed this event, we realized that understanding this concept is the first step in providing *shared understanding*, which is one critical aspect of a well-functioning BA [1].

Practitioners considered the **category scope** quite logical. However, they did not have practices targeting stakeholders' needs and terminology. They also seem unfamiliar with how the boundaries can affect the BA. When not planned or managed, artefacts tend to grow organically. As people start to use them, BAs cross several boundaries and become overloaded with multiple meanings, losing their value and decreasing trust levels [26, 1, 7, 4]. In this situation, stakeholders are not well known, which makes their needs neglected. The scope category targets those issues and directs what to consider when creating BAs, especially determining a satisfactory level of *plasticity* of the BA, which attends to stakeholders' local needs and still keeps its robustness.

The **structure category** brings new insights into principles particular to software development contexts, such as accommodating additional and experimental content [19, 4]. Practitioners already acknowledge some of the principles belonging to the structure category, such as format and means of spreading, and how considering them enhances the *ability to transfer* information among different social worlds. However, they are unfamiliar with all stakeholders' needs, focusing primarily on immediate users. Another important remark of this category is the inclusion of the principle "decide an appropriate level of content detail", which, according to the practitioners participating in this study, can reduce the chances of practitioners forking out content and creating their own artefacts.

The **management category** had a surprising and positive reaction from practitioners in both the survey and workshop, especially regarding the principle of "establish ownership of the artefact". Ownership has been explored in software engineering literature but primarily focuses on code [27, 28, 29, 30]. We emphasise that this principle, aligned with evaluation and feedback, shapes a structured strategy for managing BAs and keeping their consistency [1]. Besides, their implementation creates a favourable environment for stakeholders to feel confident about relying on BAs, as they trust the content is correct and updated.

Artefacts, in general, are recognized as necessary, but still, practitioners see them as burdensome to create and maintain [31]. Trust decreases when the produced artefact is neglected, and its use is highly affected. The results of our study have shown that practitioners perceive that the guideline contribute to creating trustworthy BAs through the three favourable factors, reliability, functionality, and predictability.

The contribution, though, differs between factors. Principles in the categories of Scope and Management strongly contribute to the BA being functional and reliable. Whilst the predictability factor has strong support from fewer and dispersed principles through the three categories, such as establishing ownership and establishing the means of spreading. These results are elucidative because, as a whole, the principles support the creation of trustworthy BAs. However, the principles can be utilized individually in specific occasions where companies, for example, want to fix a singular reliability issue and need to know which principles increase trust in BA, observing how they contribute the most to reliability.

One important note is worth mentioning. The guideline should not be seen as a process which requires all principles to be implemented in the cited order. But rather, the principles should be tailored to each BA and its context, meaning they will have different levels of importance and implemented practices.

These findings, while providing an important contribution to the software engineering literature with guideline that were inexistent by the period in which this research was concluded, also leave open questions. Would these principles remain significant in other contexts with different BA formats? Would the number of principles increase, decrease or remain the same?

We consider the developed guideline preliminary; we expect them to evolve and reveal differences as more applications are executed. Thus, new principles may be added to the guideline to match different contexts and BA, which can grow evidence in this study area.

Following our first questions, the trust factors can also differ in their contribution to trust in general. But how do they differ? Would the principles 3d. Handle scattered information, for example, have different results? We suspect that yes. Therefore, other studies could focus on identifying differences in the principles' contribution to trust when the type of BA change. The results could provide interesting insights into which principles are decisive in keeping the different BAs trustworthy.

## 6. Lessons learned

Overall, our approach was well-received by our company partner, with some initial difficulties in understanding key concepts. The most important lessons learned for us are the following:

> **Lesson #1**: Planning BAs is not an isolated activity.

The design of the BA is a group activity that involves people of different social worlds. As we gathered people that sometimes belonged within the same boundary, their different views helped to address the difficulties of what and how to represent information in the BA. Additionally, this production process allowed the solution of potentially conflicting interests [1].

> **Lesson #2**: The process of creating BAs is time-consuming.

Thinking about the structure and reconciling different perspectives requires much more than a single meeting. Usually, other stakeholders need to be involved until a consensus is reached. Initial gathering focuses mostly on brainstorming sessions where people would read the guideline and think of how to implement such principles into the artefact. As the discussions progress, the practices start to be established.

> **Lesson #3**: Start with a few boundaries and increase them as there is a need for the artefact to grow.

At the beginning of a BA creation, sometimes the boundaries do not appear straightforward, and even when they are, dealing with all of them can be complex. During the workshops, we learned that visualising how potential boundaries, e.g., one or two, could utilise already shaped directions on implementing the guideline.

> **Lesson #4**: Reviewing the stakeholders' information sources for performing tasks can facilitate outlining their needs.

Knowing the stakeholders' existing sources of information can help define the level of detail of the information added to the BA. In our workshop with BA1, some practitioners were unsure about keeping one specific piece of information as it had no owner and was updated. However, during the discussions, they found that the stakeholders could not find that information anywhere else. Therefore, they decided to keep and appoint an owner for that content and keep it up to date.

> **Lesson #5**: Mapping the guideline against a new (or existing) BA reveal preconceived assumptions.

During our workshop with BA1, practitioners realised that many assumptions were made while creating the artefact. When they confronted the few existing practices with the guideline, they exposed the lack of decisions on important properties of the BA, such as management of the BA, approach for different terminology and stakeholders' needs.

## 7. Threats to validity

In this section, we report and discuss threats to the validity of our study. We group the threat according to the classification proposed by Wohlin et al. [32]: construct, internal, external, and conclusion threats.

Construct validity refers to threats to the interpretation of theoretical constructs investigated and how they were actually investigated in the research setting. We identified one threat to practitioners' understanding of the guideline and trust factors in both workshops and surveys. During the workshop, we clarified misunderstandings and definitions of complex constructs such as boundary artefacts to minimise this threat. Besides, we avoided using the trust factors to avoid confusion or misinterpretation. Instead, we utilise questions that are straightforward to understand (see supplemental material).

Internal validity concerns the awareness of other factors that can affect the outcome of the findings other than the ones already identified. We acknowledge the existence of this threat when collecting the principles and creating the guideline. To minimise this threat, we made sure not to exclude any domains to synthesise state-of-the-art practices when creating and managing BAs.

External validity relates to the extent that the results are generalisable. To increase the generalisability of the guideline, we adopted knowledge from different areas and did not limit ourselves to the software engineering context. Yet, we acknowledge the number of participants and the type of the artefact (mostly automated) could not be enough for a significant generalisation factor. However, generalisation was not our goal; rather, it was to provide preliminary guideline that can be gradually adopted and extended. Hence, as more applications of the guideline occur, we expect the guideline to change over time to accommodate characteristics of different artefacts or contexts within software engineering. In addition, differences in practitioners' perceptions of how the principles contribute to trustworthy artefacts in other contexts.

The conclusion validity refers to how consistent the data collection allows for drawing accurate conclusions. To synthesise the literature, we applied some structure to find relevant papers. We also recorded the workshop not to miss the feedback and discussions over the principles. At last, the second and third authors joined discussions regarding designing the guideline, the structure of the workshops, and the results collected afterwards.

## 8. Conclusion

Boundary artefacts are critical for software development because they can cross several boundaries in the organisation and provide relevant input to different stakeholders. Yet, they are doomed to failure if they have enough inconsistencies that drive stakeholders not to use them, negatively affecting software projects.

In our investigation, we developed preliminary guideline that advise creating BAs. In addition, through an expert validation of the guideline, we collected practitioners' perceptions on how each guideline's principles contribute to creating trustworthy BAs. We grouped the principles collected from a literature review into three categories. The first category (Scope) focuses on the scope, displaying principles referring to defining each boundary's target audience, needs, and terminology. The second category (Structure) relates to how the artefact's content is structured to meet stakeholders' needs. The third (Management) refers to principles that can guide the establishment of practices to manage the artefact throughout time.

The expert validation revealed that the principles contribute to creating trustworthy BAs at different levels—also, the relevance of the guideline and their usefulness. Moreover, the guideline strengthen BA traits such as shared understanding, plasticity and ability to transfer. Practitioners can utilise the guideline to guide the creation or even evaluate current practices for existing BAs.

As the principles contribute differently to trustworthiness in our specific case, we believe that future application of the guideline and trust evaluation with different types of BAs, i.e. manual or automated, could provide interesting results on how the principles contribute to trust depending on the type of BA.

*Acknowledgements*

We want to thank all practitioners at Axis Communications who participated and contributed to this study. We would also like to acknowledge that this work was supported by the KKS foundation through the S.E.R.T. Research Profile project at Blekinge Institute of Technology.